\documentclass[fleqn,usenatbib]{mnras}

\usepackage{newtxtext,newtxmath}

\usepackage[T1]{fontenc}

\DeclareRobustCommand{\VAN}[3]{#2}
\let\VANthebibliography\thebibliography
\def\thebibliography{\DeclareRobustCommand{\VAN}[3]{##3}\VANthebibliography}

\usepackage{graphicx}	
\usepackage{amsmath}	
\usepackage{tabularx}

\usepackage{amsfonts}
\usepackage{graphicx}
\usepackage{subfigure}
\usepackage{mathtools}

\graphicspath{{./}{Figures6/}}

\title[Resolving period aliases]{Resolving period aliases for TESS monotransits recovered during the extended mission}

\author[Benjamin F. Cooke et al.]{
\newline
Benjamin F. Cooke,$^{1,2}$\thanks{E-mail: b.cooke@warwick.ac.uk}
Don Pollacco,$^{1,2}$
David R. Anderson,$^{1,2}$
Daniel Bayliss,$^{1,2}$
\newauthor
François Bouchy,$^{3}$
Samuel Gill,$^{1,2}$
Nolan Grieves,$^{3}$
Monika Lendl,$^{3}$
Louise D.\ Nielsen,$^{3}$
\newauthor
St\'ephane Udry$^{3}$
and
Peter J.\ Wheatley$^{1,2}$
\\
$^{1}$Department of Physics, University of Warwick, Gibbet Hill Road, Coventry CV4 7AL, UK\\
$^{2}$Centre for Exoplanets and Habitability, University of Warwick, Gibbet Hill Road, Coventry CV4 7AL, UK\\
$^{3}$Observatoire de Gen\`eve, University of Geneva, 51 ch des Maillettes, CH-1290 Veroix, Switzerland\\
}

\date{Accepted XXX. Received YYY; in original form ZZZ}

\pubyear{2020}

\begin{document}
\label{firstpage}
\pagerange{\pageref{firstpage}--\pageref{lastpage}}
\maketitle

\begin{abstract}
We set out to explore how best to mitigate the number of period aliases for a transiting TESS system with two identified transits separated by a large time period on the order of years. We simulate a realistic population of doubly transiting planets based on the observing strategy of the TESS primary and extended missions. We next simulate additional observations using photometry (NGTS) and spectroscopy (HARPS and CORALIE) and assess its impact on the period aliases of systems with two TESS transits. We find that TESS will detect around 400 exoplanets that exhibit one transit in each of the primary and extended missions. Based on the temporal coverage, each of these systems will have an average of 38 period aliases. We find that, assuming a combination of NGTS and CORALIE over observing campaigns spanning 50 days, we can find the true alias, and thus solve the period, for up to 207 of these systems with even more being solved if the observing campaigns are extended or we upgrade to HARPS over CORALIE.
\end{abstract}




\section{Introduction}
\label{sec:Introduction}

The Transiting Exoplanet Survey Satellite \citep[TESS,][]{2015JATIS...1a4003R} has recently completed its observations of the northern ecliptic hemisphere during the second half of its primary mission. As of 5 July 2020 TESS has begun its extended mission in which it will first carry out a re-observation of the southern ecliptic hemisphere, approximately 2 years after first observing it (for extended mission details see NASA 2019 senior review\footnote{\href{https://smd-prod.s3.amazonaws.com/science-pink/s3fs-public/atoms/files/SR2019_Subcommittee_Rpt.pdf}{2019 Senior Review Subcommittee Report}} and NASA response\footnote{\href{https://smd-prod.s3.amazonaws.com/science-pink/s3fs-public/atoms/files/NASA_Response_SR2019_Final.pdf}{NASA Response to the 2019 Senior Review of Operating Missions}}).

TESS has already produced numerous discoveries of planets around bright stars \citep[e.g.][]{2018ApJ...868L..39H,2018A&A...619L..10G,2019ApJ...871L..24V,2020arXiv200100952G} as well as discoveries of systems for which we observe only a single transit, \citep[e.g.][]{2019MNRAS.tmp.2805G,2019arXiv191005050L,2020MNRAS.495.2713G}, including the first monotransiting planet from TESS, \cite{2020arXiv200500006G}. The detection of monotransits is a natural consequence of the limited temporal coverage of TESS when compared to other surveys such as the Wide Angle Search for Planets \citep[WASP,][]{2006PASP..118.1407P} or Kepler/K2 \citep{2010Sci...327..977B,2014PASP..126..398H}. These monotransits have been explored via simulations \citep{2018A&A...619A.175C,2019AJ....157...84V}, which show we can expect a large number of monotransit candidates in the TESS primary mission data. These predictions have been verified by recent monotransit searches including \citet{2020MNRAS.498.1726M} who find 15 southern hemisphere candidates and the NGTS monotransit team (see \citet{2020arXiv200500006G,2020MNRAS.495.2713G} for a description of the pipeline used) who have identified over 50 candidates from the first few TESS sectors alone. Many of these candidates exhibit additional extended mission transit events as discussed later in this paper. Because of the close dependence of monotransit detections on survey coverage the determination of orbital periods for single-transit candidates will be heavily influenced by the continuation of TESS into its extended mission \citep{2019A&A...631A..83C}.

With the data from the first year of the extended TESS mission (i.e. Year 3 of TESS), most southern ecliptic TESS targets will be observed for about twice as long. However, these additional observations will be carried out approximately 2 years after the initial observations. Due to this, there will be significant gaps in the photometric baseline which may hide additional transits. This effect is most pronounced for those systems which only transited once during the primary mission and will only transit once more in the first year of the extended mission. Based on only two transits we cannot infer the true orbital period, we know only that it is bounded. The maximum period is given by the separation of the two observed transits which is on the order of $\sim$2\,years. The minimum period is given by the constraint that only one transit is seen in each TESS observing run, approximately 10\,days for a single sector of observations. Therefore we are left with a discrete set of period aliases \citep{2019A&A...631A..83C}.

Period aliases are an important consideration when analysing non-continuous observations. This is a common problem for ground based observations due to the daytime and weather interruptions in the data; any signal short enough to fit within a day can be missed and it requires sufficient night time observations to rule out these potential periods. However, for longer period systems, this is a significant issue for space based observations as well \citep{2019AJ....157...19B,2010ApJ...722..937D}. Some attempts have been made to use period priors to prioritise aliases to search \citep{2019arXiv191204287D} but this is far from foolproof. To actually rule out aliases requires additional data in the form of photometry and/or spectroscopy. This paper looks at exactly what effect additional observations have on the number of period aliases of a given system and how the effect changes depending on the amount and quality of these observations. 
We explore both photometry and spectroscopy by simulating each method, and investigating how this reduces the set of period aliases for TESS discoveries. For a recent comparison of photometric and spectroscopic follow-up methods see \cite{2020MNRAS.495..734C}.

This paper is laid out in the following manner. Section \ref{sec:Observations} discusses the stellar and planetary population used as well as the simulated TESS, photometric and spectroscopic data. Section \ref{sec:Alias determination} details our attempts to use the data to identify and analyse the period aliases. Section \ref{sec:Results} presents the results of our simulation and analysis while Section \ref{sec:Conclusions} outlines our conclusions.

\section{Observations}
\label{sec:Observations}

\subsection{Simulation population}



To predict the number of planets for which TESS will observe two individual transits, we first create a planetary population. TESS observations of this population are then simulated. To do this we proceed as in \cite{2018A&A...619A.175C} which we summarise below.

We use the TESS Input Catalogue Candidate Target List version 8 \citep{2019AJ....158..138S} accessible via the Mikulski Archive for Space Telescopes (MAST\footnote{\href{http://archive.stsci.edu/tess/tic\_ctl.html}{http://archive.stsci.edu/tess/tic\_ctl.html}}) as our stellar population, which includes Gaia DR2 parameters \citep{2018A&A...616A..10G}. We perform the following simple cuts on this population based on TESS-band magnitude ($3.0 \leq m_{\rm T} \leq 17.0$) and effective stellar temperature ($2285 \leq T_{\rm eff} \leq 10050$\,K) leaving 4789372 targets in the southern ecliptic hemisphere.

We generate planets around these stars by drawing from period-radius bins under particular occurrence rates given by \cite{2015ApJ...807...45D} for M-stars and \cite{2019AJ....158..109H} for FGK-stars. Note that this is an updated set of occurrence rates compared to the procedure laid out in \cite{2018A&A...619A.175C}. In particular, \cite{2019AJ....158..109H} uses the final Kepler Data Release \citep[DR25,][]{2016ksci.rept....3T} and the improved stellar parameters made available via GAIA DR2 \citep{2018A&A...616A..10G}. More specific transit parameters are then calculated using equations and distributions presented in \cite{2010arXiv1001.2010W} and \cite{2018ApJS..239....2B}. We include a uniform distribution in $\cos{i}$ for inclination used to calculate transit probability, but assume all planets around the same star share an inclination value.

\subsection{TESS observations}

To identify which systems would be observed to transit once in both the primary and extended TESS missions, we simulate TESS observations. Since it will be the first hemisphere to be re-observed in the extended mission, and due to our use of southern ecliptic hemisphere based follow-up observatories, we only simulate southern ecliptic hemisphere TESS targets. 
Our methods follow that described in \cite{2019A&A...631A..83C}. We simulate observations only at times for which there are 
data points not affected by systematics from the TESS primary mission. This has the effect of producing a more realistic coverage fraction per sector and accounts for systematic effects such as momentum dumps. We assume that the primary mission coverage can be replicated during the extended mission, noting that in the extended mission Full Frame Images will be taken with a improved cadence of 10 minutes.


Based on our simulated TESS coverage we can now determine which planet transits are detectable with TESS. Using our planet parameters and a 5$\rm^{th}$ order polynomial TESS noise approximation \cite{2018AJ....156..102S} we calculate a Signal-to-Noise ratio, $S/N$ for each planet. Those planets which exhibit a transit during a region of TESS observations that manage to achieve the required $S/N \geq 7.3$ are recorded as detectable. For more specific details regarding the simulated TESS observations see \cite{2019A&A...631A..83C}.





Once the simulation is complete we select all those planets with one detectable transit during the TESS primary mission and one detectable transit during the first year of the extended mission as our double transit sample.

\subsection{Simulated photometry}
\label{sec:Simulated photometry}

We simulate additional photometric data using the Next Generation Transit Survey \citep[NGTS,][]{2018MNRAS.475.4476W}. For this additional photometry we follow the procedure being employed by NGTS \citep[see][]{2019MNRAS.tmp.2805G,2019arXiv191005050L}. Observations are simulated from sunset to sunrise using a cadence of 12\,s. We use the NGTS noise model as a function of host magnitude shown in Figure \ref{fig:noise} and compare this to the signal size of the planet, $(R_p/R_\star)^2$. Additionally, we allow the use of multiple NGTS telescopes if required to reach the necessary $S/N$ level. To calculate the $S/N$ level for multiple telescope we use a $1/\sqrt{N}$ relation as has been shown to be a good approximation by \cite{2020MNRAS.494.5872B}. Where NGTS can achieve a $S/N\geq 3.0$ we use the simulated photometry. Below this $S/N$ threshold we assume that photometry is of insufficient quality to help constrain the period of the system and do not simulate follow-up photometry.

To best utilise photometric time we only simulate observations on nights when we predict a transit may occur (see Sect. \ref{sec:Time taken vs nights observed} for details including accounting for transit time uncertainties). Since this paper explores those systems with transits in both the primary and extended TESS missions we can use the TESS transit times and coverage to create a set of discrete period aliases and we therefore only simulate NGTS observations on nights that at least one alias predicts will contain a transit. This is a realistic strategy that avoids wasted time that would occur by observing on nights which provide no period constraining information.

We can then run this strategy for a fixed number of nights, observing some subset of those. In the main body of this paper we consider an overall time-span of 50 nights. Appendix A then includes the results for a range of different maximum time-spans (\ref{tab:N_aliases}). As an additional measure, we include telescope downtime from instrumental and weather effects. This is included by assuming that a random 20\% of nights are unavailable for observations (for more details see \citealt{2020MNRAS.494..736C}).

\subsection{Simulated spectroscopy}

\begin{figure}
    {\includegraphics[width=\columnwidth]{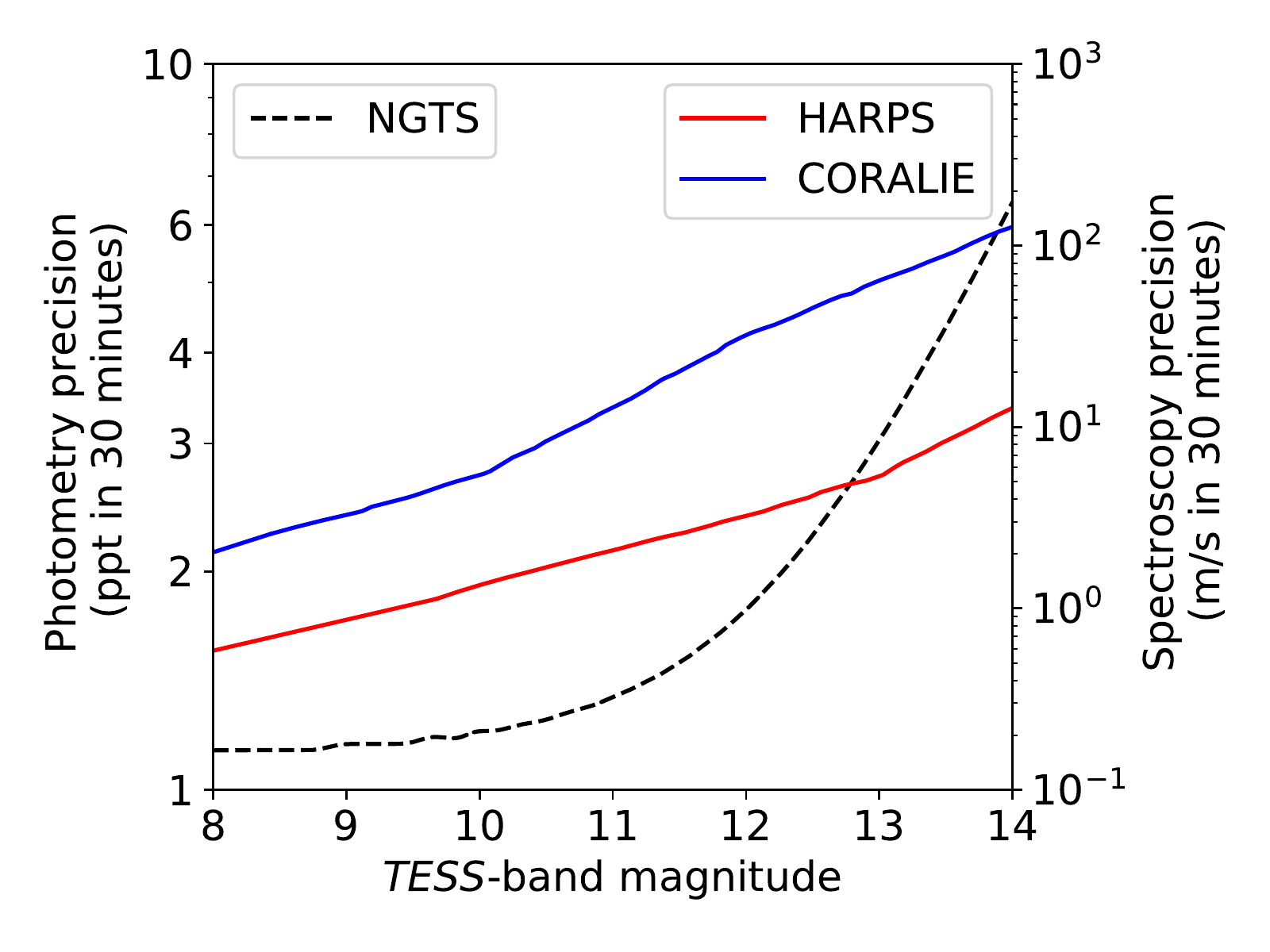}}
    \caption{Instrumental precision as a function of target TESS magnitude for NGTS, HARPS and CORALIE (adapted from Figure 14 of \protect\citealt{2018MNRAS.475.4476W} and Figure 10 of \protect\citealt{2017MNRAS.465.3379G}).}
\label{fig:noise}
\end{figure}

Spectroscopic data is simulated using the High-Accuracy Radial-velocity Planetary Search \citep[HARPS,][]{2003Msngr.114...20M} on the ESO 3.6\,m telescope and CORALIE \citep{2000A&A...354...99Q} on the Euler 1.2\,m telescope. We simulate an observing strategy of one 30 minute exposure every 7 days for each candidate, including a noise value based on the noise models of the instruments as a function of host magnitude shown in Figure \ref{fig:noise}. The amplitude of the radial velocity signal is a function of the true period (which is known), the stellar mass (which is known), and the planetary mass (which is unknown). Therefore we first predict the planetary mass using the planetary radius and the MRExo mass-radius relation\footnote{\href{https://github.com/shbhuk/mrexo}{https://github.com/shbhuk/mrexo}} from \cite{2019arXiv190300042K}. For each radius value we draw a random value from a mass distribution generated using MRExo.

As with photometry, this procedure can be run for any time-span, the difference being that now the number of data points is one seventh of the number of nights. In this paper we again consider an overall time-span of 50 nights using CORALIE with alternative results (including using HARPS) presented in Appendix A. As with the simulated photometry in Section \ref{sec:Simulated photometry}, we account for telescope downtime from instrumental and weather effects as well as scheduling conflicts. This is included by assuming that a random 20\% of nights are unavailable for observations (for more details see \citealt{2020MNRAS.494..736C}).

\section{Alias determination}
\label{sec:Alias determination}

\subsection{Period aliases from TESS}

Based on our 
simulated TESS observations in Years 1 and 3 (primary and extended mission) we select the subset of planets which exhibit one transit during each year. The separation of these transit times gives an upper limit to the period of the planet as well as the knowledge that the true period must be equal to this separation divided by an integer. Assuming that the minimum period that a planet could have and still only be observed once in a sector is $\sim10$\,days we can make a list of all possible period aliases. Some of these periods however can be ruled out by the TESS coverage we already have. We test each period alias by comparing with the TESS coverage. For each alias we create a list of all times at which the system would transit (including times before the Year 1 transit and after the Year 3 transit) and check if there are any TESS observations within half a transit duration of any of these. If there are TESS observations that overlap with any transit time (excluding the two discovery transits) we can rule out that alias as one that would result in an additional transit which has not been seen. In this way we reduce the list of period aliases down to only those allowed based on the TESS photometry.

\subsection{Period aliases using additional data}

With the newly simulated photometry and spectroscopy we can now attempt to better constrain the true period of each monotransit and rule out more aliases. We do this first using the TESS and follow-up photometric data, then the TESS and spectroscopic data, then finally using the TESS, photometric and spectroscopic data.

To combine the NGTS data we proceed on a nightly basis. For each night we check if any of the possible period aliases predict a transit to occur. If so, we simulate observations. If no transit is seen it shows that the true period is not one of those that predicted a transit this night and we therefore remove all periods that predicted a transit that night from our list of aliases. If a transit is seen we can use this third transit to remove any aliases that do not predict this additional detection. This process repeats each night until the period is solved or we reach our observing limit.

Ruling out aliases using spectroscopic data requires a different approach. Firstly, we require a minimum of 5 radial velocity measurements before we begin to rule out aliases. Once we have reached this threshold every time we simulate a new spectroscopic data point we fold the radial velocity data on each alias and plot in phase space. 
For each allowed period we model the radial velocity variation assuming a circular orbit and fitting for the the amplitude, phase, and gamma velocity. We then calculate the residuals between our simulated measurements and this fit. If the average of the absolute residuals is greater than 1.5 times the noise threshold at the chosen magnitude we reject the alias. Otherwise the alias is still a valid option for the system. Since we require at least 5 data points the average of the residuals is unlikely to reach this threshold for the true period. In multiple runs the true period was never rejected. 
This process is repeated until the period is solved or we reach our observing limit.

To find the viable aliases using a combination of TESS photometry, photometric follow-up, and spectroscopic follow-up, we simply examine the alias sets from TESS and photometry and from TESS and spectroscopy and allow only those aliases present in both sets. This is carried on on a nightly basis to determine the earliest point at which a system is solved.

\section{Results and Discussion}
\label{sec:Results}

\subsection{TESS extended mission monotransits}

Based on our simulated observations we expect 759 monotransits will be found in the TESS southern ecliptic primary mission. Of these, approximately 65\% will transit again in the first year of the extended mission with most only transiting once more. The result is that slightly over 50\% of the primary mission southern monotransits will transit once more in the first year of the extended mission, giving a total of 395 planets that will exhibit one detectable transit in both the primary and extended TESS missions for the southern ecliptic hemisphere. 
This number exceeds those presented in \cite{2018A&A...619A.175C} and \cite{2019A&A...631A..83C} by a factor of 2-3. This is due to the updated occurrence rate used in this study. The occurrence rate used here comes from \cite{2019AJ....158..109H} which uses a more up to date and larger sample based on the Kepler mission. Therefore these results will be more robust than similar studies based on outdated occurrence rates.

\subsection[test title1]{Number of aliases, $N_{\rm aliases}$}

For each southern ecliptic hemisphere planet that TESS will observe to transit once in each year we have produced a list of period aliases. From this list we determine the number of aliases, $N_{\rm aliases}$ for each system based on different subsets of observational data. For each system we then calculate $N_{\rm aliases}$ using four different combinations of observational data; TESS data only, TESS and additional photometry, TESS and additional spectroscopy and finally TESS, and additional photometry and spectroscopy. We define the number of aliases found using TESS data alone as $N_{\rm aliases,i}$ and the updated value found by taking into account different combinations of follow-up data as $N_{\rm aliases,f}$.

\begin{figure*}
\centering
    {\includegraphics[width=0.8\linewidth]{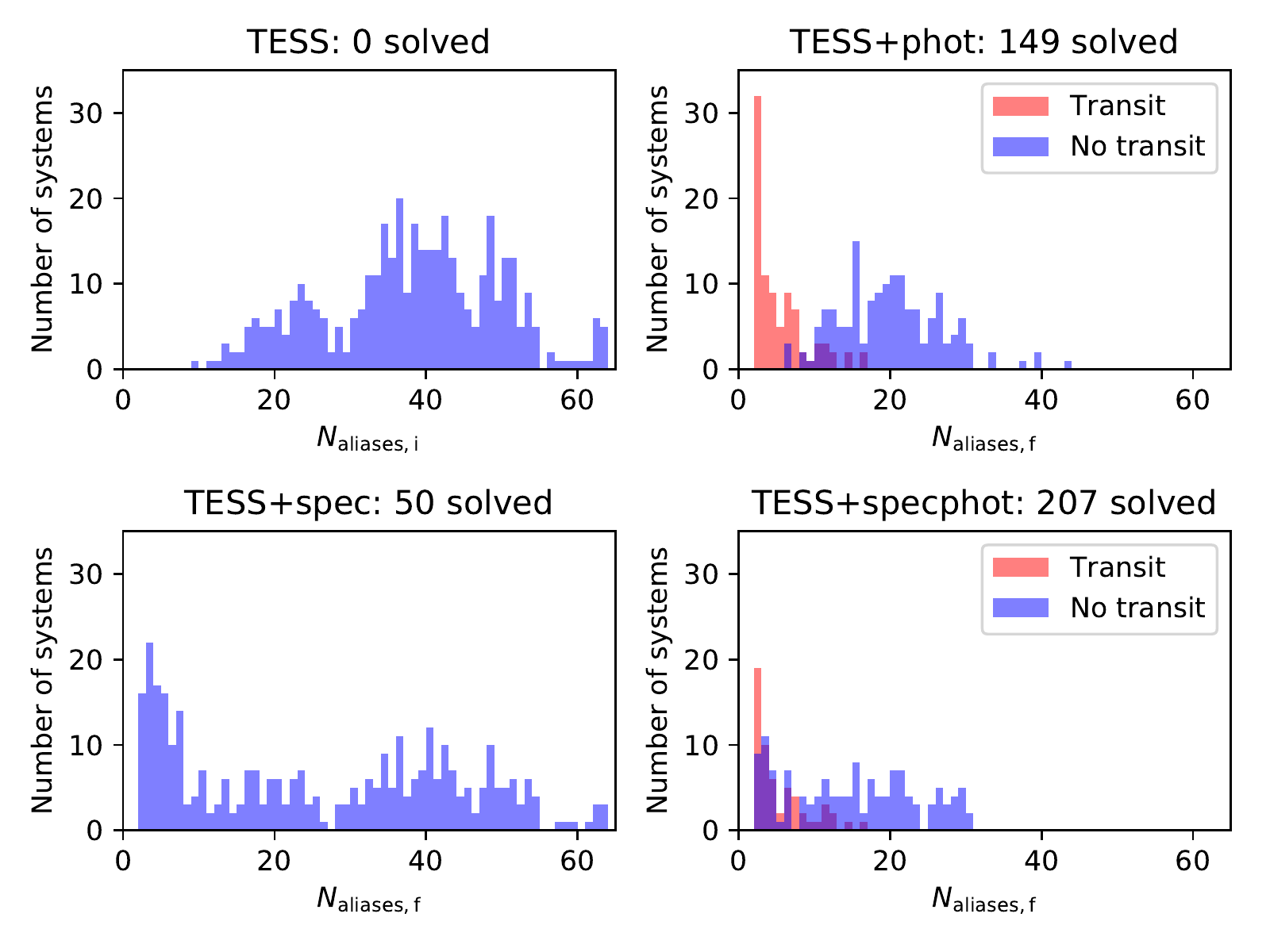}}
    \caption{Distributions of the number of period aliases from each combination of observational data. The four plots show $N_{\rm aliases,i}$ and the three $N_{\rm aliases,f}$ values for the four combinations of observations. For the right hand column plots (those including additional photometry) the data is split into those for which an additional transit is seen (red) and those for which no additional transit is seen (blue). Each plot tile denotes the data used and how many systems have been solved with this data (these systems are not shown in the plots to avoid overwhelming the histograms). Here we assume a time-span of 50 days and use CORALIE for spectroscopy.}
\label{fig:N_aliases}
\end{figure*}

Figure \ref{fig:N_aliases} shows the distribution of the number of aliases from TESS, $N_{\rm aliases,i}$, and the number of aliases from the three combinations of follow-up data, $N_{\rm aliases,f}$.

For each plot in Figure \ref{fig:N_aliases} we show the value of $N_{\rm aliases,i}$ or $N_{\rm aliases,f}$ for all unsolved systems when accounting for the four regimes of observational data. The solved systems (those for which the period is known, i.e. $N_{\rm aliases,f}=1$) are not shown in the plots as they show large spikes at 1 overwhelming the rest of the distributions. The number of solved systems, and therefore the number of systems not shown, is in the title of each plot. The top left plot shows the distribution of the number of aliases per system using TESS data alone, $N_{\rm aliases,i}$. This gives an average value of 38 aliases per system, however it should be noted that there is a double peak in the distribution. This corresponds to those systems which are observed in multiple TESS sectors. Systems with more TESS coverage have correspondingly fewer aliases leading to a secondary peak around 20. Systems with a single TESS sector per year have more aliases and peak around 40.

\begin{figure}
    \begin{subfigure}[Transit depth distribution
    \label{fig:depth_dist}]
    {\includegraphics[width=\columnwidth]{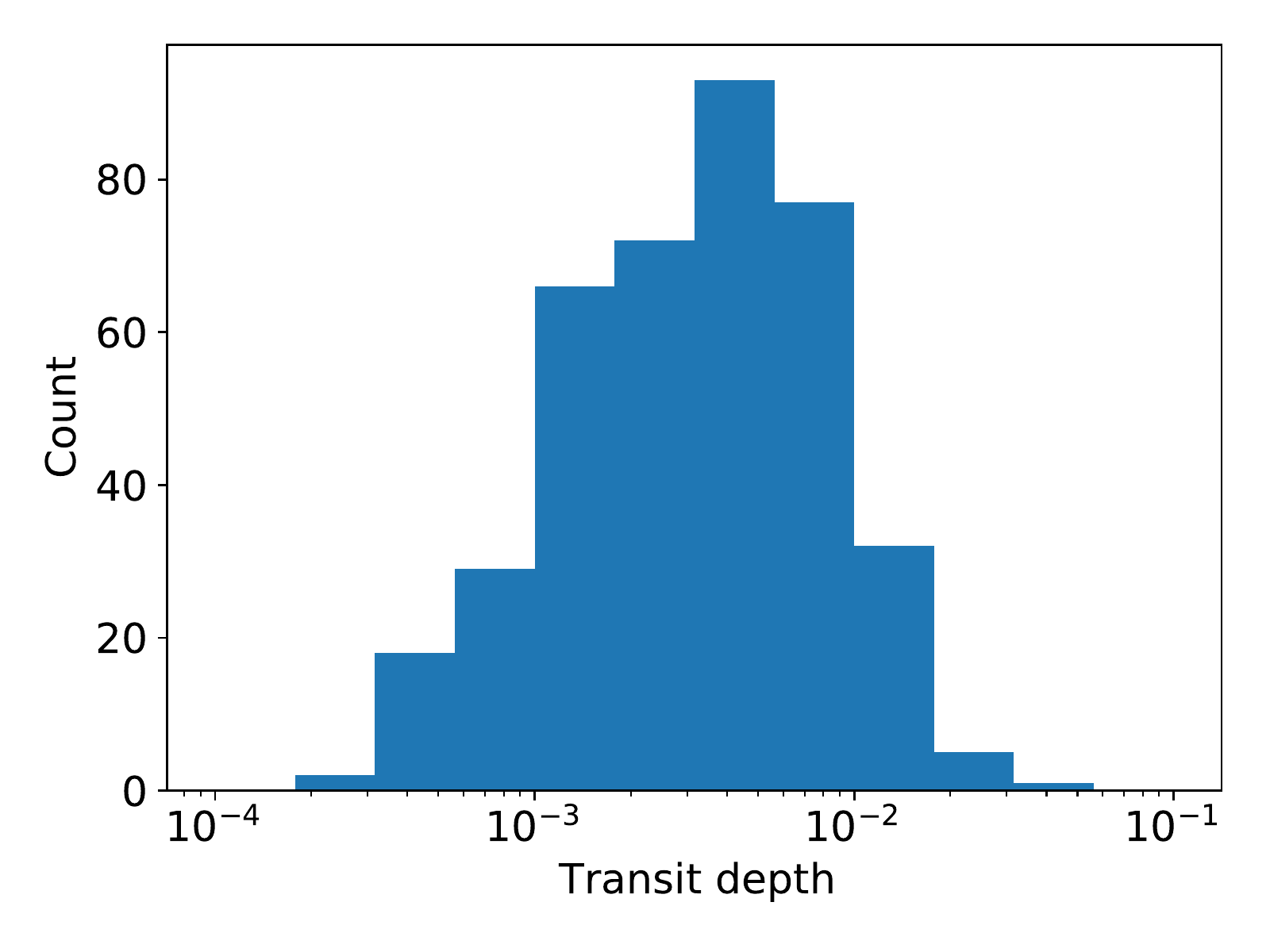}}
    \end{subfigure}
    \begin{subfigure}[RV semi-amplitude distribution
    \label{fig:K_dist}]
    {\includegraphics[width=\columnwidth]{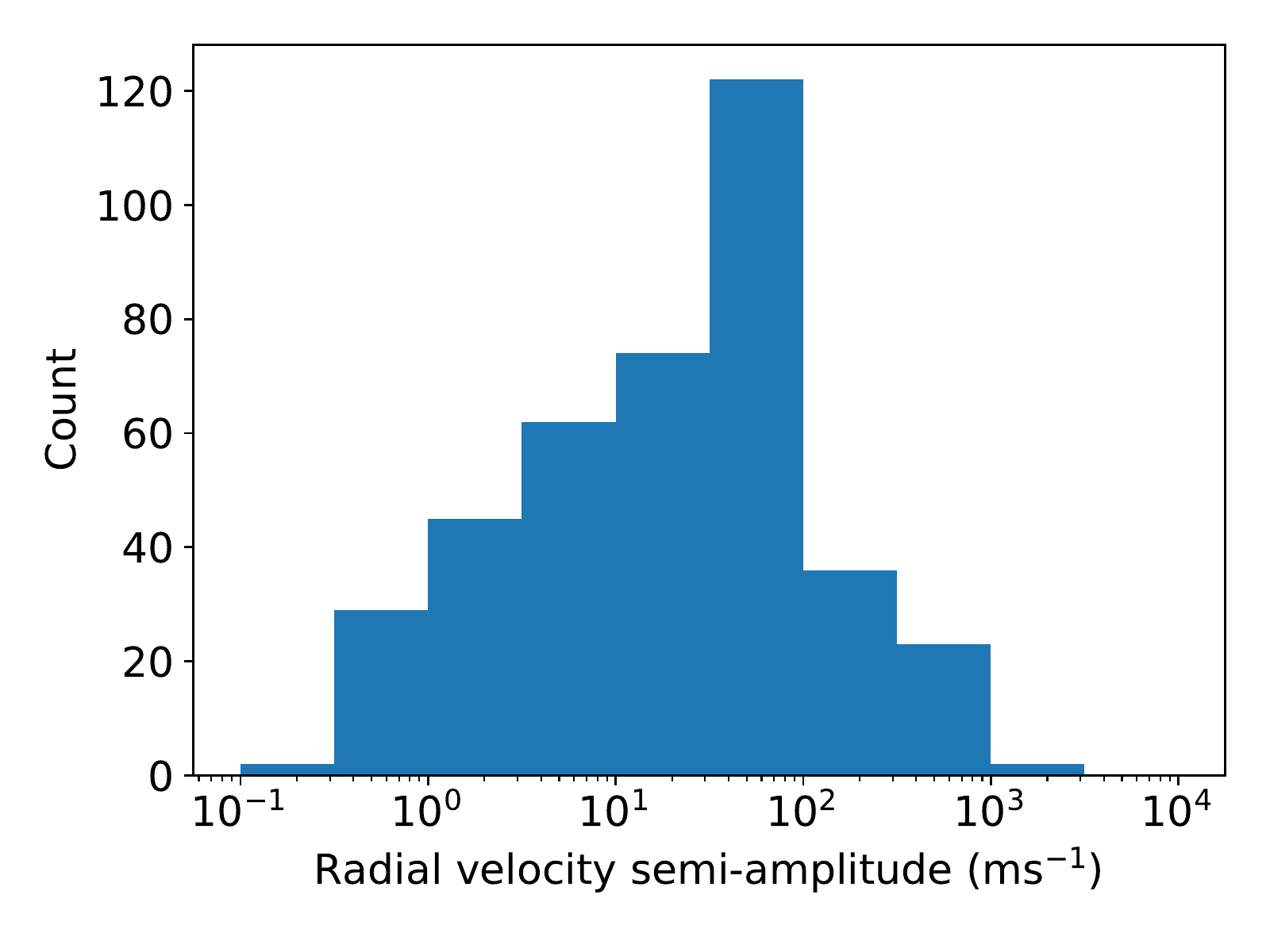}}
    \end{subfigure}
    \caption{Distribution of signal size for photometry and spectroscopy for all considered monotransits.}
\label{fig:signal_dist}
\end{figure}

The top right and bottom plots show the distributions of $N_{\rm aliases,f}$ based on three combinations of follow-up data. As can be seen photometry solves significantly more systems than spectroscopy (149 and 50 systems solved respectively) but combining the two methods leads to a number of solved systems greater than simply adding the uniquely solved systems together (207 solved systems). Figures including additional photometry, those in the right hand column, are split depending on whether the photometry caught additional (one or more) transits. It is seen that catching a transit leads to a large reduction in $N_{\rm aliases,f}$ (see \citealt{2020arXiv200500006G} for an example of this in practice) and all of the systems solved from additional photometry alone (149) had at least one additional transit detected. Included in the plots are those systems for which the photometry is of insufficient $S/N$. Transits may occur during observations of these systems but would not be detected with sufficient confidence. For these systems additional photometry cannot help to constrain the period. Additionally, some of the systems result in spectroscopic signals smaller than the respective noise level. These points are present in the plot but additional spectroscopy would not be able to rule out any periods. Figure \ref{fig:signal_dist} shows the distributions of both transit depth and radial velocity semi-amplitude for the monotransits considered. For both photometry and spectroscopy we require $S/N \geq 3$ to simulate observations.

As well as the number of solved systems proved by each combination of observational data these distributions show the number of aliases for those unsolved systems. For TESS photometry only, all 395 systems are unsolved and the average (mean) value of $N_{\rm aliases,i}$ is $\sim$\,38. When photometry is included we have 246 unsolved systems with an average $N_{\rm aliases,f}$ value of $\sim$\,14. Using spectroscopy instead gives 345 unsolved systems with an average $N_{\rm aliases,f}$ value of $\sim$\,26. Finally, including both photometric and spectroscopic data on top of the TESS photometry gives 188 unsolved systems with an average $N_{\rm aliases,f}$ value of only $\sim$\,12.

From Figure \ref{fig:N_aliases} we see that the number of systems solved when combining the follow-up photometry and spectroscopy is larger that the sum of the systems solved using each method individually. Further analysing this we find that of the systems solved by spectroscopy, 19, or 38\%, are also solved using photometry. This absolute value is obviously then the same for the photometrically solved systems also solved by spectroscopy but in this case it is only 13\% of the total.

Taking this analysis forward to the systems solved when combining the two methods we find that 27 of these (13\%) are not solved by either photometry or spectroscopy individually. These 27 systems are therefore systems for which each method manages to rule out different subsets of aliases but is incapable of completely solving the period. Taking the common aliases left from both follow-up methods results in these systems being solved. It should also be noted that photometric and spectroscopic data provide complementary information about the radius and mass of the planet, another reason that a combination of follow-up data is highly desirable.

\begin{figure}
\centering
    {\includegraphics[width=\columnwidth]{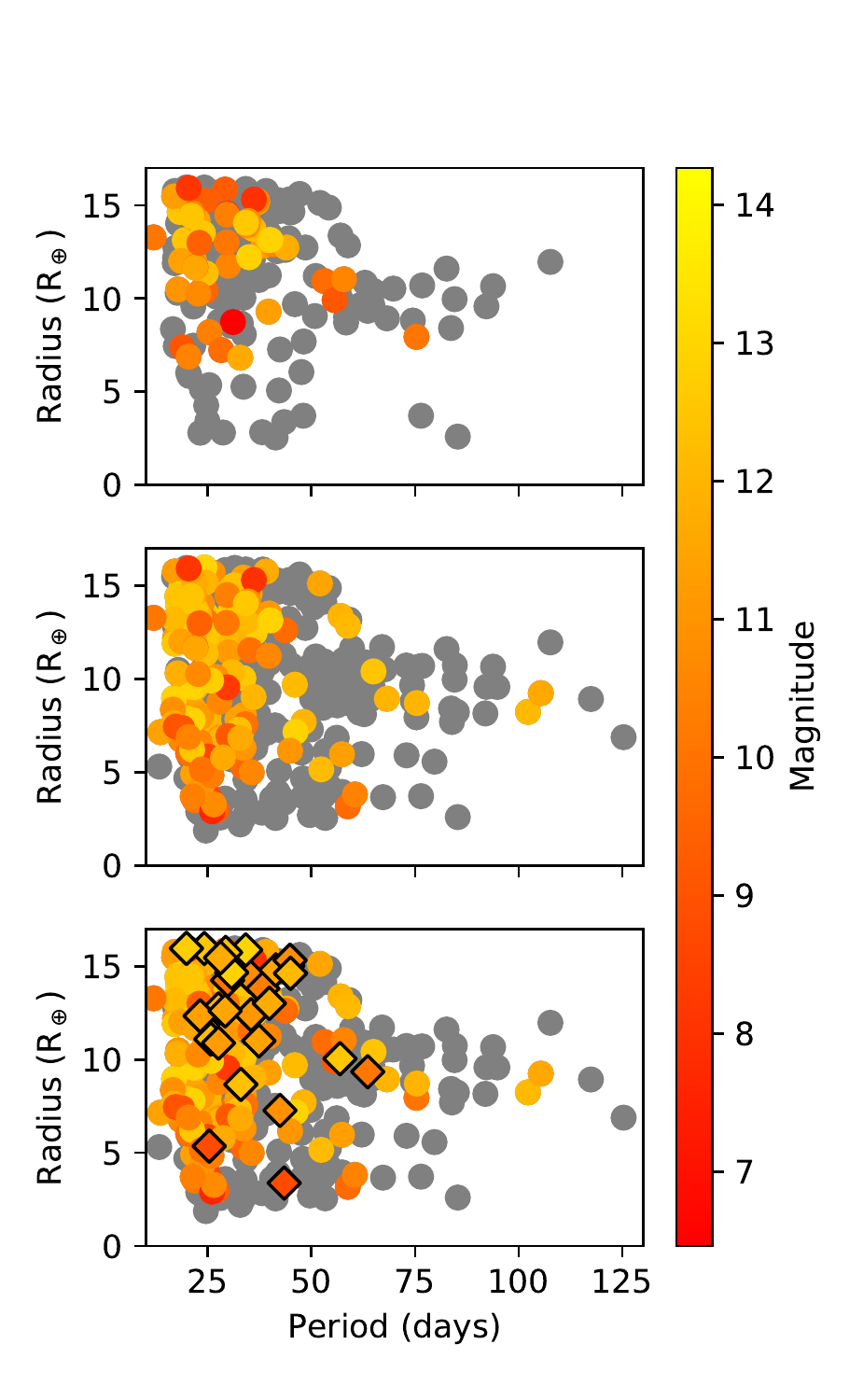}}
    \caption{Period and radius distribution of solved systems. The top plot shows the distribution for spectroscopically solved systems, the middle is that for photometrically solved systems and the bottom shows the combination of both methods. Diamond points denote systems only solved when both methods are combined. The colour of each point denotes its TESS-band magnitude.}
\label{fig:period_radius}
\end{figure}

Figure \ref{fig:period_radius} shows the distribution of solved systems with orbital period and planetary radius. 
The top panel shows the results for spectroscopy with systems solved being coloured by magnitude and unsolved systems being grey. Systems with insufficient $S/N$ are omitted from the plot. The middle panel shows the corresponding plot for photometry and the bottom panel shows the combination of the two methods. In this panel all systems are plotted regardless of either photometric or spectroscopic $S/N$ and systems that are only solved when the methods are combined are shown as diamonds.

From this plot we see that the majority of systems solved by spectroscopy have radii between 10 and 15$R_\oplus$. This is as expected since predictions and observations suggest this radii range corresponds to the most massive planets which are the ones that spectroscopy is most sensitive to. The photometric planets are more spread in radius since they tend to be insensitive to radius (and mass) as long as the transit depth is above the sensitivity threshold. Both methods favour shorter periods since these are obviously easier to confirm within a given observing program. It also seems that the subset of planets solved only when both data sets are combined (diamonds in Figure \ref{fig:period_radius}) have a broader period distribution. Once again, this is unsurprising as the additional data help to confirm these more challenging systems.



\subsection[test title2]{True period index, $I$}

\begin{figure}
    {\includegraphics[width=\columnwidth]{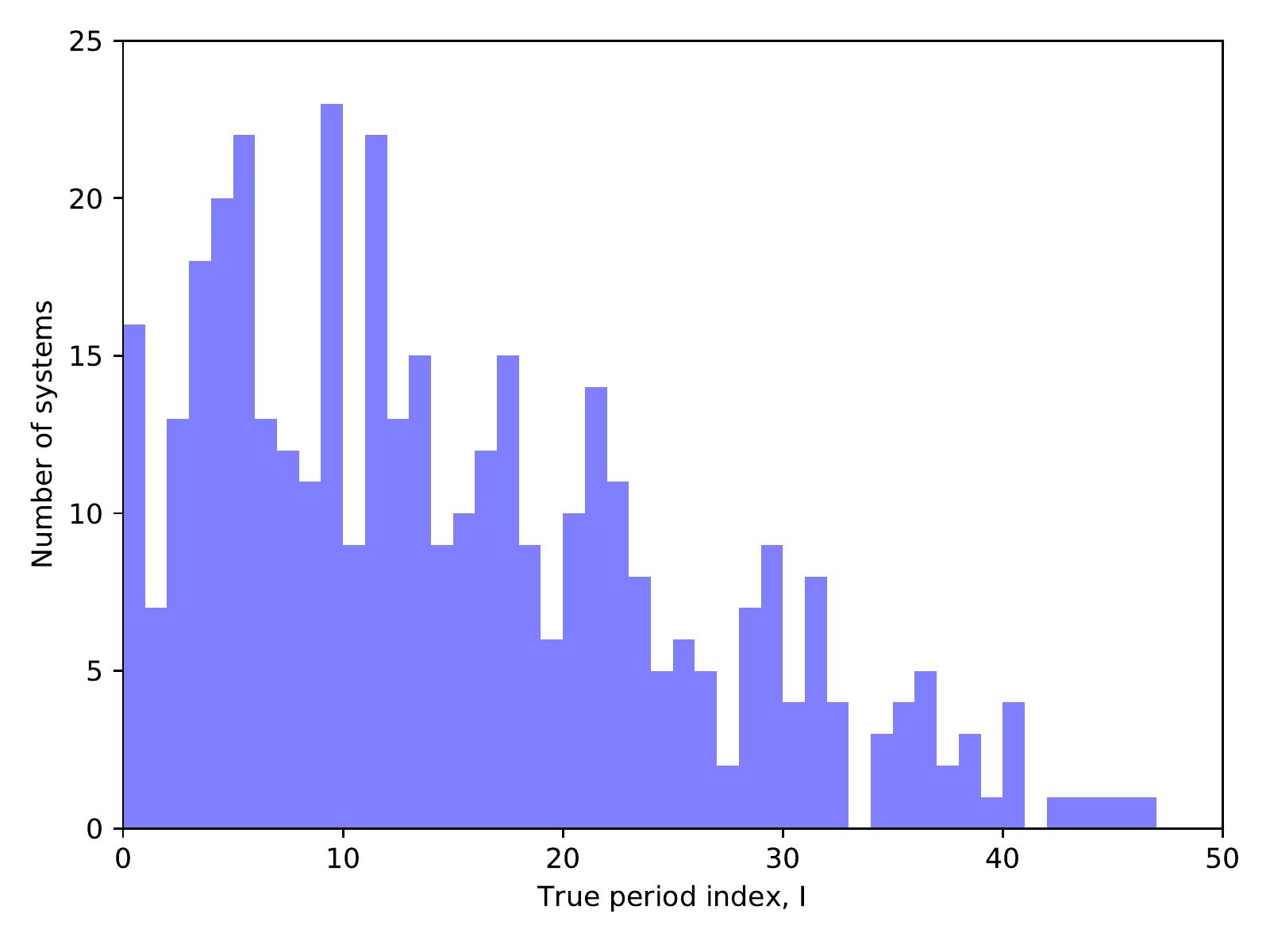}}
    \caption{True period index, $I$ for all 395 TESS systems which show a single transit in each of the the Year 1 and Year 3 data sets. A value of $I=0$ means the true period is the smallest allowed alias.}
\label{fig:I_all}
\end{figure}

Based on the increased likelihood of detecting short period planets (transit probability scales as $P^{-\frac{2}{3}}$) it might be natural to assume that, even for unsolved systems, the true period is most commonly the shortest of the period aliases. However, this is frequently untrue. Figure \ref{fig:I_all} shows the distribution of the true period index $I$ for all 395 systems that transit once in each of the TESS primary and extended missions. $I$ is the index of the true period in the list of all period aliases based on the TESS data alone. A value of $I=0$ means that the true period is the shortest allowed alias.

Figure \ref{fig:I_all} shows that the average value of $I$ is $\sim$\,15 and only 16 of 395 systems have $I=0$. This strongly supports the need for the additional photometric and/or spectroscopic observations discussed in this paper. Even working up from the shortest periods (i.e. lowest $I$ values) it still requires an average of $\sim$\,14 aliases to be ruled out before the true period is found. Using additional observations to reduce $N_{\rm aliases,i}$ and therefore reduce $I$ is vital. There have been some attempts at predicting the true period of a systems from a set of aliases using statistical methods \citep{2019arXiv191204287D} but, due to the statistical nature, the reliability for individual system is poor thus an attempt to reduce $N_{\rm aliases,i}$ is still needed.

Another method for identifying the true period is to use the transit shape. Some attempts have been made to use a Bayesian inference based on the transit shape and an assumption of zero eccentricity to constrain orbital period to $\sim$\,10\% from a single K2 transit \citep{2016MNRAS.457.2273O}. Longer period systems however are more likely to have non-zero eccentricity \citep{2005A&A...431.1129H,2011MNRAS.414.1278P} therefore this method is not robust.
Additionally, many of these systems will be discovered in the TESS 30\,min cadence images \citep{2018A&A...619A.175C,2019A&A...631A..83C} where the transit shape is less well defined.

\subsection[test title3]{Improvements in number of aliases, $\Delta N_{\rm aliases}$}

\begin{figure*}
\centering
    {\includegraphics[width=0.8\linewidth]{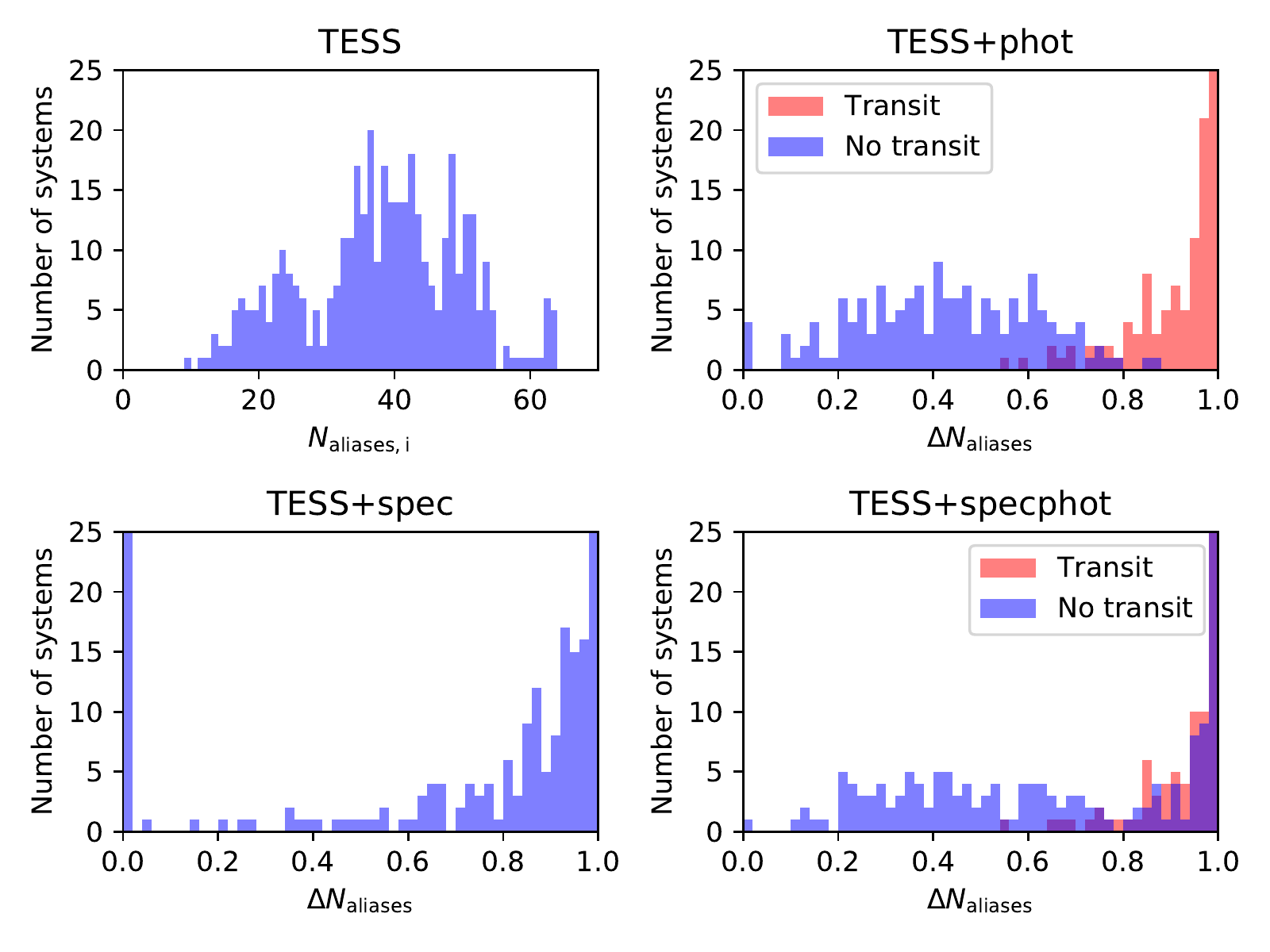}}
    \caption{Improvements in number of period aliases found when additional observational data is accounted for. Shown are distributions of $\Delta N_{\rm aliases}$ for the three combinations of additional observations. The top left plot repeats the top left plot in Figure \ref{fig:N_aliases} to show the values of $N_{\rm aliases,i}$ by which the other three subplots are divided. A value of $\Delta N_{\rm aliases}=0$ means no improvement has been made whereas $\Delta N_{\rm aliases}=1$ means the system has been solved. As in Figure \ref{fig:N_aliases} the right hand column plots (those including additional photometry) are split into those systems for which an additional transit is seen (red) and those for which no additional transit is seen (blue). Here we assume a time-span of 50 days and use CORALIE for spectroscopy. The spikes at 0.0 and 1.0 overwhelm the rest of the data and are truncated here, the values are given in the text.}
\label{fig:Delta_NN}
\end{figure*}

As well as exploring the absolute number of aliases as a result of different amounts and methods of observations it is useful to explore how these values change. This is shown in Figure \ref{fig:Delta_NN} which plots $\Delta N_{\rm aliases}$ where $\Delta N_{\rm aliases}$ is defined as

\begin{equation}
    \Delta N_{\rm aliases} = \frac{N_{\rm aliases,f}-N_{\rm aliases,i}}{N_{\rm aliases,i}-1}.
\end{equation}
\noindent
In this formalism a value of $\Delta N_{\rm aliases}=0$ means that $N_{\rm aliases,i}$ has not been improved by additional follow-up data whereas a value of $\Delta N_{\rm aliases}=1$ means all incorrect aliases have been ruled out and the period of the system has been solved. The value can be interpreted as the fraction of aliases that each method can rule out.


To better show the change in the number of allowed period aliases these plots show all 395 systems, including both solved and unsolved systems. This leads to 2 peaks in each plot at 0.0 and 1.0. The peak at 1.0 corresponds to the solved systems, i.e. $N_{\rm aliases,f}=1$. The peak at 0.0 corresponds to those systems where $N_{\rm aliases,f}=N_{\rm aliases,i}$. For photometry these are the systems for which NGTS cannot reach sufficient $S/N$ meaning the data cannot be used to detect transits and for spectroscopy these systems are those for which the spectrograph noise (CORALIE in these plots) is comparable to, or greater than, the RV signal of the planets meaning the period cannot be seen under the scatter. These peaks overwhelm the rest of the data and are truncated in the plots but are described fully below.

Once again we see a clear separation in the right hand column plots based on the detection of an additional transit. Those systems seen to transit again are almost all above 0.8 (meaning over 80\% of the aliases are eliminated) with 149 being solved fully. For systems without an additional transit the fraction lies mainly between 0.1 and 0.7, the full distribution gives 297 systems ($\sim$\,75\%) that have more than half their aliases eliminated. Ignoring the unsolvable systems (insufficient $S/N$) this is $\sim$\,76\%.

Spectroscopy shows a slightly different distribution. The peak at 1.0 is much smaller than photometry (50 compared to 149) but the peak at 0.0 is larger (210 compared to 4). Exterior to the two peaks the spectroscopy distribution is then geared towards higher fractions with a larger clump of systems between 0.9 and 1.0. In fact, we see that almost all systems with sufficient $S/N$ have more than half of their aliases ruled out (93\%). 
This is as expected since spectroscopy data continues to add information on each subsequent night whereas photometry data is less useful until catching an additional transit. The reason that comparatively few spectroscopy systems are able to go from only a few remaining aliases to completely solved ($N_{\rm aliases,f}=1$) is that by this point only similar period aliases are left. This method of spectroscopic observations is efficient at ruling out significantly different periods but finds it very difficult to identify the correct period between two similar values. Photometry is less affected by this feature as an additional transit is, in general, very good as discriminating between two similar periods.

Once again, combining the two methods gives the best result with the overall improvement in fraction of aliases eliminated by spectroscopy being further improved by systems with additional transit from photometry. When using both methods only 1 out of 395 systems are not improved and 335 systems ($\sim$\,85\%) have more than half of their period aliases eliminated.

\subsection{Time taken vs nights observed}
\label{sec:Time taken vs nights observed}

\begin{figure}
    \begin{subfigure}[Photometric hours observed
    \label{fig:nights_observed_phot}]
    {\includegraphics[width=\columnwidth]{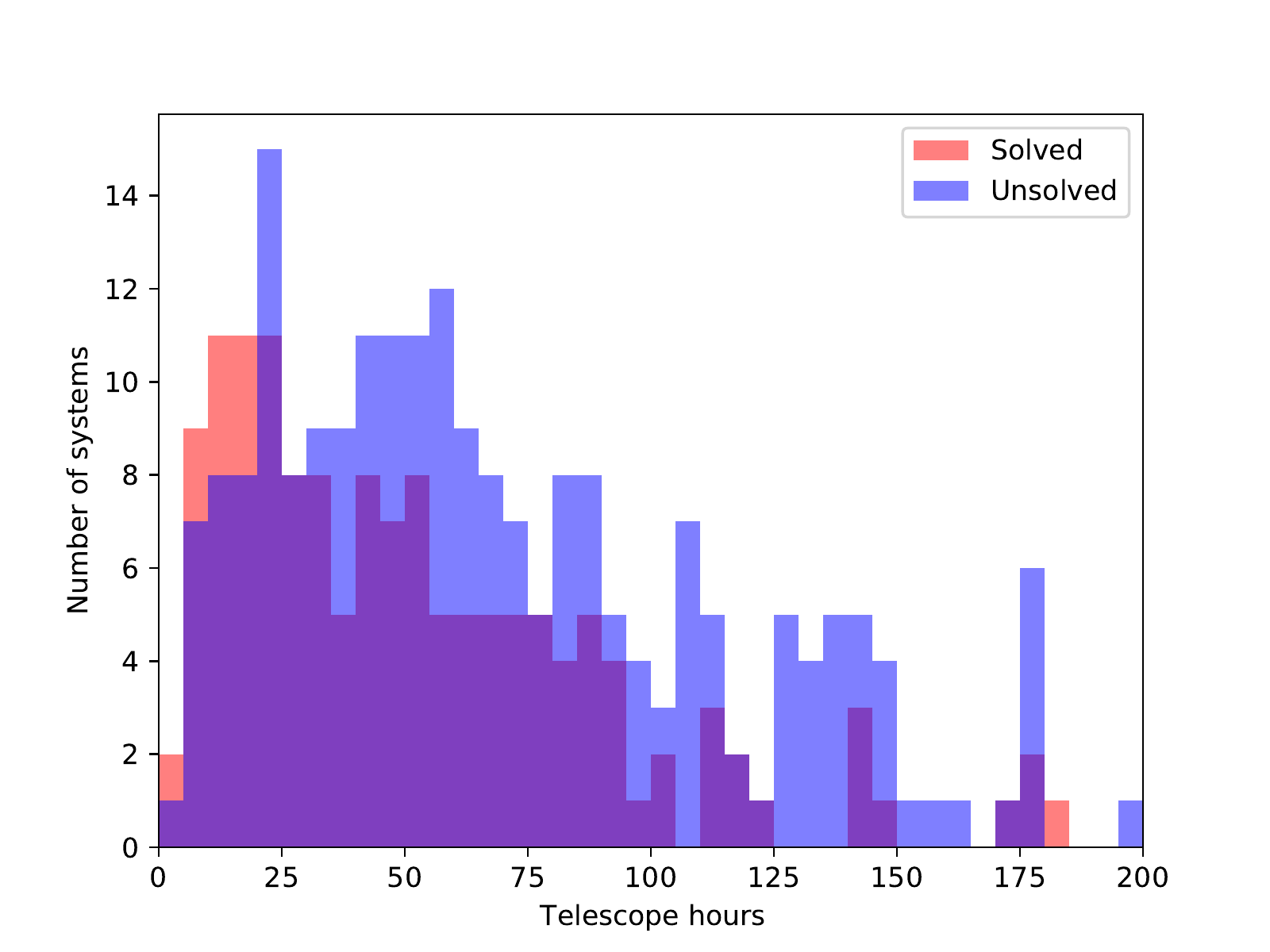}}
    \end{subfigure}
    \begin{subfigure}[Spectroscopic hours observed
    \label{fig:nights_observed_spec}]
    {\includegraphics[width=\columnwidth]{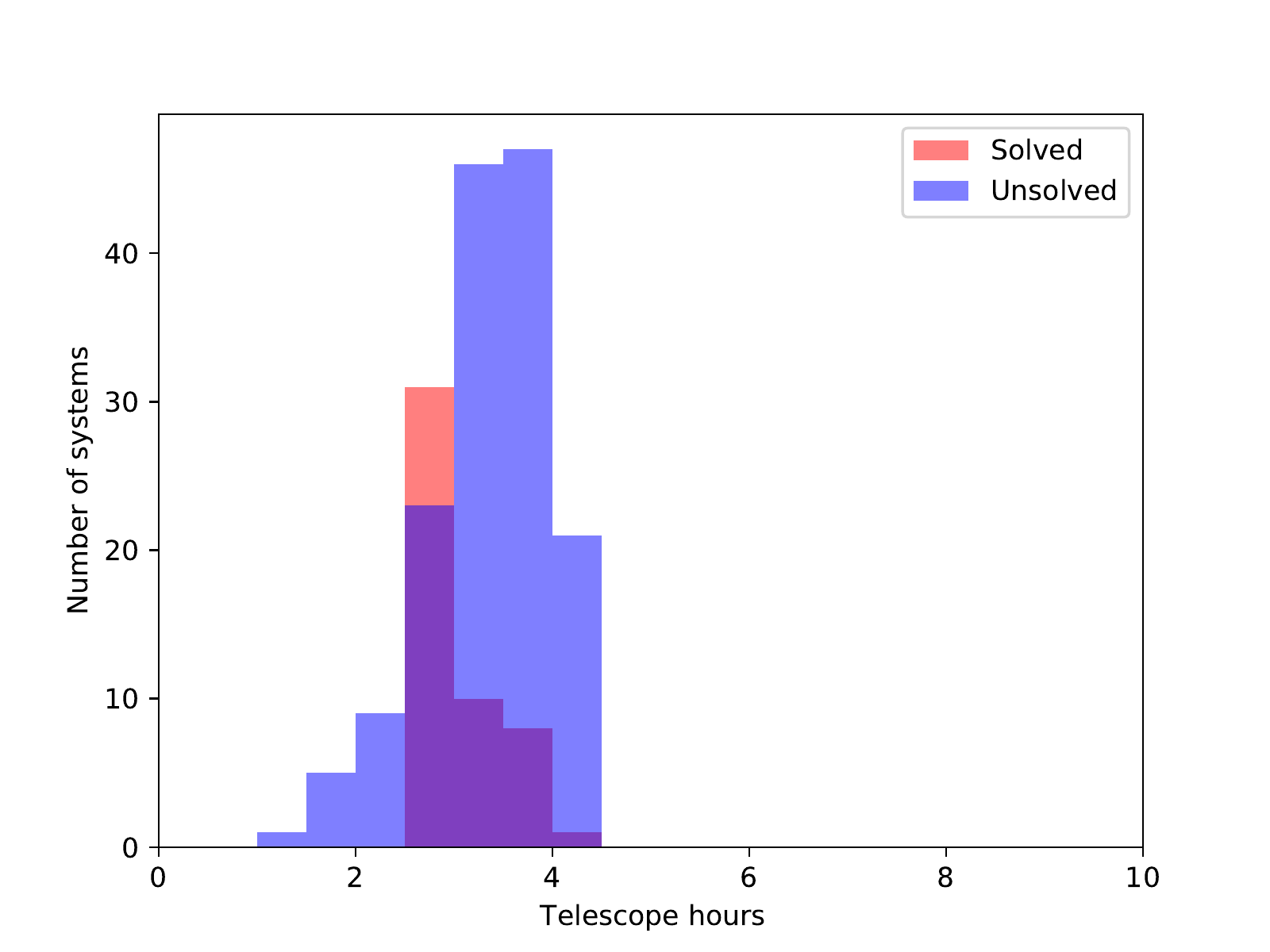}}
    \end{subfigure}
    \caption{Telescope hours utilised by each system for photometry and spectroscopy. Systems are coloured for solved (red) and unsolved (blue).}
\label{fig:nights_observed}
\end{figure}

An additional aspect of this simulation that must be considered is that though our simulation is run for 50 days no system is observed for that many nights. Firstly, each system is only observed by each method until that method has managed to solve the system after which further observations are not simulated. Secondly, for spectroscopy, a system is only observed once every seven days, allowing for weather. This means that the number of nights for which a system is observed spectroscopically is approximately equal to $\frac{T}{7}\times0.8$ where $T$ is either the time until the system is solved or the time-span of the simulation, whichever is smaller. For photometry a system is only observed when one of its period aliases predicts a transit will occur. This means that the actual number of nights observed can be a small fraction of the time over which the simulation is run. Additionally, we do not necessarily observe for entire nights at a time. For spectroscopy we use a noise model that assumes half-hour exposures so each simulated night of observing is only half an hour of spectroscopy time. For photometry we observe each night for the full length of any transit that is predicted to occur that night i.e. if only a partial transit is predicted we only simulate observing time during that window. As such the total number of telescope hours is different from the number of nights observed and we show the values here. Figure \ref{fig:nights_observed} shows plots of the number of hours that systems are observed for over the 50 days for which this simulation is run.

Additionally, for photometry we build in a buffer time based on the propagation of transit timing uncertainties from the TESS transits (as a bonus this buffer time will allow us to obtain out of transit baseline photometry, helping to improve transit characterisation). For each period alias the uncertainty is equal to $2\sigma_t/N$ where $\sigma_t$ is the uncertainty on the transit centre time of each of the TESS transits and $N$ is the integer value used to calculate each period alias. We choose $\sigma_t = 1\rm hr$ for this calculation, twice the cadence for the TESS full Frame images which are the source of the majority of our targets. Propagating this uncertainty forward we find the uncertainty on successive transit times, $\sigma_T$, to be given by

\begin{equation}
    \sigma_T = 2\sigma_t(\frac{1}{2}+\frac{n}{N}),
\end{equation}
\noindent
where $n$ is the number of periods after the second TESS transit. For this we assume that our simulated observations begin approximately 100 days after the end of the TESS extended mission for the southern ecliptic hemisphere. This is likely an overestimation since we should be able to start following up targets from early TESS sectors while the extended mission is ongoing. For each estimated transit time for which we observe, our observation window is extended by this value to ensure transits are not missed due to inaccuracies in the TESS transit timing measurements.

In these figures we see that each system is observed for a significantly smaller number of hours than are available within the observing span of 50 nights. Each method is split into 2 categories: those which are solved ($N_{\rm aliases,f}=1$, red) and those which remain unsolved ($N_{\rm aliases,f}>1$, blue). Systems with insufficient $S/N$ are not shown. 
For photometry the average number of hours observed is 62.8 and 114.2 for the solved and unsolved systems respectively (ignoring those systems for which we have insufficient $S/N$ that are not observed at all). The use of multiple telescopes is accounted for when calculating telescope hours. 
This highlights the efficacy of a targeted photometric campaign as opposed to a stare campaign, where we simply wait for an additional transit. For spectroscopy we see the number of hours observed is 2.8 and 3.1 for solved and unsolved systems respectively. Based on the strategy of one half-hour every seven days, combined with the weather constraints we expect an average unsolved value of approximately $\sim$3.3 which matches what is seen.

Looking at these plots combined shows us that the majority of solved systems are solved with the first 5 hours of spectroscopy time or the first 80 hours of photometry time. Therefore, if attempting to most efficiently solve the largest number of systems, the ideal strategy may be to observe each system for 10 nights (assuming 0.5 hours per night for spectroscopy and an average of 8 hours per night for photometry) and if not solved move on to a different system. However, this neglects the fact that some systems are of greater scientific interest than others. Therefore, it actually requires a more careful consideration of individual system parameters to determine which systems may be worth a larger fraction of available telescope time.



\subsection{Effect of increasing follow-up time}

\begin{figure}
    {\includegraphics[width=\columnwidth]{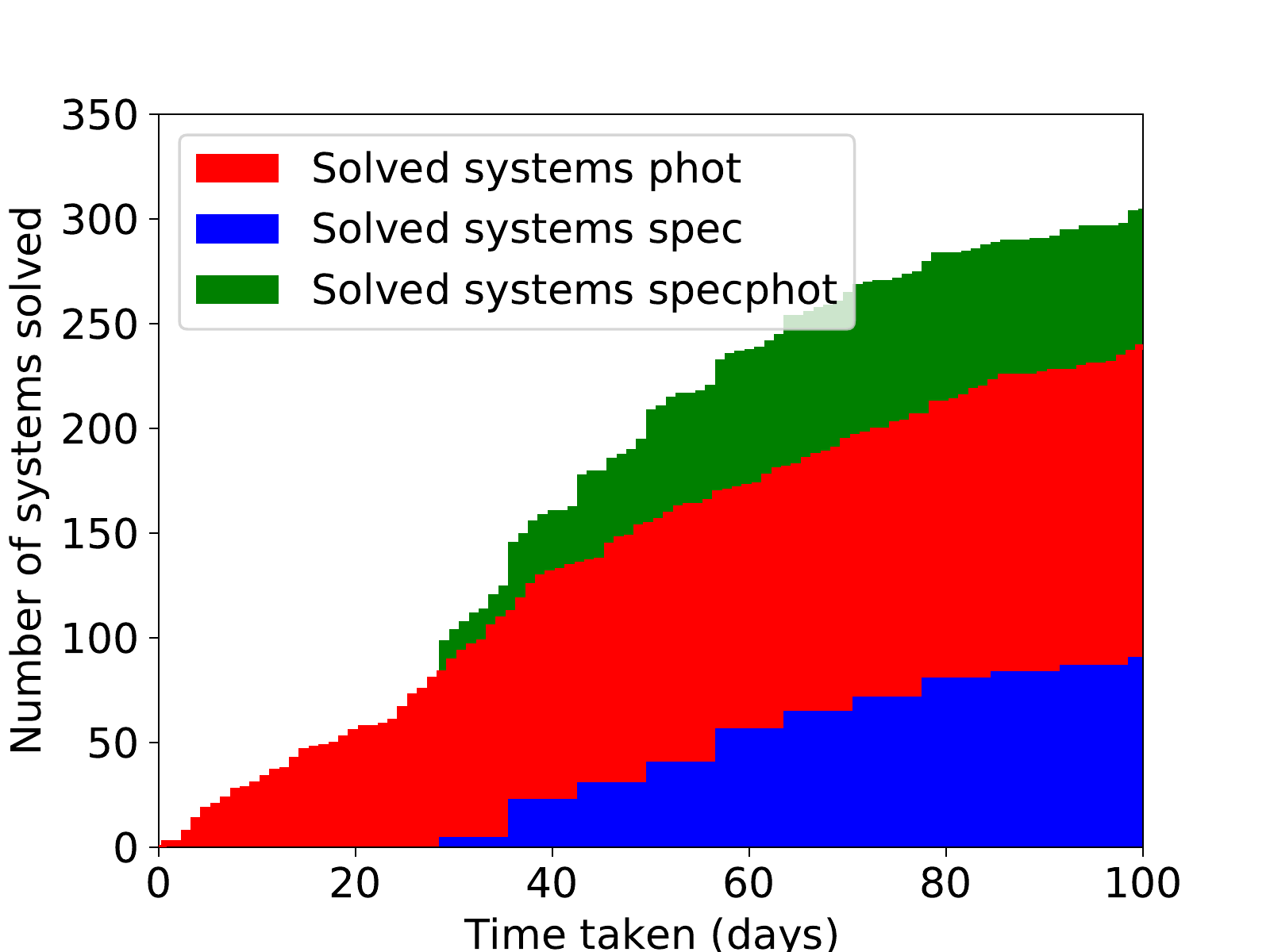}}
    \caption{Cumulative histogram of solved systems as a function of additional photometric or spectroscopic time. The distribution runs for 100 days with photometry being carried out by NGTS and spectroscopy consisting of one CORALIE point every seven days. Only systems which reach $N_{\rm aliases,f}=1$ within 100 days are shown.
    }
\label{fig:Solved_hist}
\end{figure}

The plots and analysis so far have used a time-span of 50 days and CORALIE for spectroscopy. Table \ref{tab:N_aliases} shows how the $N_{\rm aliases,f}$ values change for different combinations of time-span and spectroscopy instrument. However, it is also interesting to look in more detail at how the number of solved systems increases as the time-span is allowed to increase. This is shown in Figure \ref{fig:Solved_hist}.

Figure \ref{fig:Solved_hist} shows that the fraction of systems that can be solved using either photometry or spectroscopy is a function which will asymptote before reaching unity. The reason for this is as explained above, systems with sufficiently small transit or RV signals will become lost in the noise of their respective instrumental precision's and will thus never be solved. 

Initially we see that the number of systems solved through photometry rises quite slowly. This is because, in the early stages of the follow-up campaign, the coverage is sparse enough that only catching an additional transit will solve a system fully. Also, because there has not been time for other aliases to be ruled out by non-detections, only systems for which the observed transits produce no aliases will solve a system. As the observations continue the rate of solving begins to increase. This is due to the fact that at this stage, we are beginning to see enough aliases ruled out through nights of non-detections that catching a third transit is more likely to remove all aliases that are left. Additionally, the likelihood of a fourth transit being seen increases which dramatically reduces the number of aliases.

Spectroscopically solved systems only begin to show up around 28 days since we require at least 5 data points, taken once every 7 days (including a data point at time zero). As we see in Figure \ref{fig:nights_observed_spec} there is a spike in solved systems upon reaching this threshold, however, due to the randomness of weather effects different systems reach this threshold on different nights with 28 simply being the earliest possible time. The slow increase from here on is then the result of the difficulty of using spectroscopy to fully solve a system. Were we to focus on systems for which spectroscopy can rule out 80\% of aliases this peak would be much higher but it is challenging for spectroscopy to definitively decide between similar periods.

The number of systems solved by combining the two methods is always above those solved using either method alone as expected. However, as time increases the fractional addition in number of solved systems caused by this combination above those solved by photometry alone is reduced. As seen above, at 50 days the total number of systems is increased from 149 to 207, 39\%, but by 100 days the increase is from 240 to 305, only 27\%. This seems to be due to the fact that by this time, the number of systems solved by spectroscopy is increasing only very slowly, therefore spectroscopy is adding little to the number of overall solved systems whereas the systems solved via photometry are still increasing, albeit slower than they were originally, due to the shorter and easier systems having been solved already.

It is also of interest to note how many transits are required for a photometric solution. It is seen that for the majority of systems (122/149) only one additional transit is required to fully solve the period. The remaining solved systems are done so with two additional transits. This matches what we see for photometry in Figure \ref{fig:Solved_hist}. The initial quick increase in solved systems is a result of those systems that can be solved by catching only one more transit. This then levels off as we approach the longer period systems which require a larger number of additional transits to solve.

\section{Conclusions}
\label{sec:Conclusions}

Catching two widely separated transits of a long period transiting system is insufficient for identifying the true period of the orbit. We have taken the TESS primary and extended mission observations of the southern ecliptic hemisphere as an example and showed that there are approximately 395 exoplanets with a single transit detected in each year. Based on the TESS data alone we show this leaves an average of $\sim$38 period aliases per system with the true alias being, on average, the 15$^{\rm th}$ smallest possibility. 
Building on this, we have simulated attempts to mitigate these aliases using additional photometric and spectroscopic data.

We find that the amount of additional photometric and/or spectroscopic observations taken can have a significant impact on both the number of fully resolved systems (defined here as systems with $N_{\rm aliases,f}=1$) and the mean number of aliases for the unsolved systems. The results presented in Figures \ref{fig:N_aliases} though \ref{fig:Solved_hist} and Table \ref{tab:N_aliases} can be summed up by the following points.

\begin{enumerate}

\item For all simulation lengths photometry fully solves the period of significantly more systems than spectroscopy; roughly three times more for a 50 day simulation (Figures \ref{fig:N_aliases} and \ref{fig:Solved_hist}).
\item Combining both methods results in the highest total number of solved systems. At 50 days 13\% of these were not solved by either method independently (Figure \ref{fig:N_aliases}).
\item When selecting only those systems which reach the required $S/N$ threshold for each method photometry rules out at least half of the period aliases for 76\% whereas the corresponding value for spectroscopy is 93\% (Figure \ref{fig:Delta_NN}).
\item Spectroscopically solved systems favour higher mass planets which leads to a peak between 10 and 15$R_\oplus$ which corresponds to the largest mass planets. Additionally it is easier to solve shorter period systems (Figure \ref{fig:period_radius}).
\item Photometrically solved systems are insensitive to mass or radius assuming they exceed the $S/N$ threshold. As for spectroscopy, shorter period systems are easier to solve (Figure \ref{fig:period_radius}).
\item Systems solved through a combination of the two methods tend to have a broader period distribution, extending to longer period systems (Figure \ref{fig:period_radius}).
\item For an observing period of 50 days the majority of systems solved by photometry are done so in under 80 hours of telescope time. All spectroscopically solved systems are solved in under 5 hours (Figure \ref{fig:nights_observed}).
\item We suggest that, with the singular goal of confirming the period for the greatest number of systems, a 50 day follow-up campaign is preferred when accounting for the cost-benefit ratio. However, if attempting to characterise more scientifically interesting systems different campaign lengths may be more efficient.
\item Longer observing periods lead to more solved systems but both methods asymptote before solving all systems (Figure \ref{fig:Solved_hist}).
\item Using HARPS instead of CORALIE leads to an increase in the number of spectroscopically solved systems by a factor of 3-5 which decreases as the observing period increases (Table \ref{tab:N_aliases}).

\end{enumerate}



\section*{acknowledgements}

BFC acknowledges a departmental scholarship from the University of Warwick. DLP acknowledges support from the Royal Society during this period.

The contributions at the University of Warwick by PJW, DRA, and SG have been supported by STFC through consolidated grants ST/L000733/1 and ST/P000495/1.

We thank the referee for their helpful comments.


\section*{Data availability}
The data underlying this article will be shared on reasonable request to the corresponding author.



\bibliographystyle{mnras}
\bibliography{Period_alias_final.bib}




\appendix

\section{Number of aliases}
\label{sec:Number of aliases}

\begin{table*}
\begin{minipage}{\textwidth}
\centering
\caption{$N_{\rm aliases,f}$ results from 10 combinations of time-span and spectroscopic instrument. For each combination of observations we show the number of solved systems. $N_{\rm aliases,f}$ then refers to the mean number of aliases for all unsolved systems. The bold row matches the data plotted in Figures \ref{fig:N_aliases}, \ref{fig:Delta_NN} and \ref{fig:nights_observed}. Data for the TESS only simulation are not shown as they are unchanged by follow-up strategies.}
\label{tab:N_aliases}
\begin{tabular}{cc|cc|cc|cc}


Time-span & Spectroscopy & Solved & $N_{\rm aliases,f}$ & Solved & $N_{\rm aliases,f}$ & Solved & $N_{\rm aliases,f}$ \\
(days) & instrument & (phot) & (phot) & (spec) & (spec) & (specphot) & (specphot) \\
\hline
30 & CORALIE & 86 & 19.2 & 9 & 33.7 & 104 & 18.4 \\
30 & HARPS & 90 & 18.5 & 41 & 31.9 & 129 & 16.6 \\
40 & CORALIE & 118 & 16.8 & 21 & 29.4 & 156 & 15.8 \\
40 & HARPS & 131 & 16.7 & 94 & 23.4 & 220 & 14.1 \\
\bf{50} & \bf{CORALIE} & \bf{149} & \bf{14.2} & \bf{50} & \bf{25.7} & \bf{207} & \bf{11.6} \\
50 & HARPS & 150 & 13.7 & 191 & 17.9 & 298 & 8.8 \\
60 & CORALIE & 170 & 11.9 & 57 & 22.5 & 242 & 9.6 \\
60 & HARPS & 180 & 12.8 & 232 & 17.4 & 323 & 7.9 \\
70 & CORALIE & 187 & 10.8 & 82 & 24.9 & 254 & 10.2 \\
70 & HARPS & 198 & 10.9 & 243 & 17.5 & 340 & 7.5 \\

\end{tabular}
\end{minipage}
\end{table*}


\bsp	
\label{lastpage}
\end{document}